\documentstyle[aps,epsfig,twocolumn,prl]{revtex}
\begin{document}
\title{Phase transitions in nonequilibrium $d$-dimensional models with $q$ absorbing states}
\author{Adam  Lipowski$^{1),2)}$ and Michel
Droz$^{1)}$}
\address{
$^{1)}$ Department of Physics, University of Gen\`eve, CH 1211
Gen\`eve 4, Switzerland\\
$^{2)}$ Department of Physics, A. Mickiewicz 
University,61-614 Poznan, 
Poland
}
\date{\today}
\maketitle
\begin{abstract}

A nonequilibrium Potts-like model with $q$ absorbing  states is studied
using Monte Carlo simulations.  In two dimensions and $q=3$ the model
exhibits a discontinuous transition.  For the three-dimensional case
and $q=2$ the model exhibits a continuous, transition with $\beta=1$
(mean-field).  Simulations are inconclusive, however, in the
two-dimensional case for $q=2$.  We suggest that in this case the model
is close to or at the crossing point of lines separating three
different types of phase transitions.  The proposed phase diagram in
the $(q,d)$ plane is very similar to that of the equilibrium Potts
model.  In addition, our simulations confirm field-theory prediction
that in two dimensions a branching-annihilating random walk model
without parity conservation belongs to the directed percolation
universality class.
\end{abstract}
\section{Introduction}

One of the main achievements of statistical physics during the past decades is
the understanding of the universal properties of systems near an equilibrium 
second order phase transition. 
Universality classes, characterized by a small number of parameters allow us 
to understand why different systems have the same critical properties~\cite{RG}.

One important model in equilibrium statistical physics is the so-called
q-states Potts model~\cite{WU}, which can be used to describe a large class of
physical systems. Its rich critical behaviour is to a large extent
understood. It is known that the important parameters which determine
its critical behaviour are its dimensionality  $d$ and the degeneracy
of its ground state $q$.

The relatively complete understanding of equilibrium phase transitions
has yet no nonequilibrium counterpart.  However, it is becoming
evident that some analogies between equilibrium and nonequilibrium
systems could be made.

For example, a classification into universality classes is particularly
evident for one-dimensional nonequilibrium models with absorbing
states~\cite{HAYE1}, which exhibit a phase transition between active
and absorbing phase in their stationary state.  A prime example of a
universality class is the directed percolation one, which is typical to
models with a single absorbing state~\cite{DP}.  The
so-called  parity-conserving (PC) universality class is typical for
models with certain conservation laws or
symmetries~\cite{PC,TT,LIP96,HAYE97}.

However, it is necessary to investigate nonequilibrium models in higher 
dimension to have a complete characterization of the possible universal 
properties.

Analytical approach to high dimensional problems is based mainly on
field-theory methods. Recently, interesting results were obtained along
this line by Cardy and T\"auber~\cite{TAUBER} who clarified the role of
parity conservation and also discovered some new universality classes.
However, this technique is applicable  only to certain particle
systems~\cite{MUNOZ} and a large class of two- and higher-dimensional
models with absorbing states cannot be treated within such a method.
Accordingly, our understanding of higher dimensional nonequilibrium
models is rather limited.  As for continuous phase transitions some
models with single absorbing state were shown to fall into the DP
universality class~\cite{CONTACT}.  Results were also reported
for a two-dimensional model with two absorbing states~\cite{HAYE97}.
These models are often motivated by problems of surface catalysis~\cite{COMM2}.
In some models with absorbing states discontinuous phase transitions are also 
known to occur~\cite{DICKMAN,HAYE1,INFINITE}. 

A subclass of models with absorbing states are the so-called
branching annihilating random walk (BARW) ones.  In BARW models each
particle can react (annihilate, branch, diffuse, etc.) according to
prescribed rules.  It turns out that BARW models for $d>1$ are much
different from the above mentioned (surface-catalysis) models.  
Some of the results
obtained from field-theory method~\cite{TAUBER} have been confirmed
using numerical methods.  For example Szab\'o and Santos confirmed the
existence of logarithmic corrections in two-dimensional
parity-conserving BARW models~\cite{SZABO}.  However, for parity
non-conserving particle systems Monte Carlo simulations~\cite{TT} seem
to be in disagreement with field-theory results~\cite{TAUBER}.

A characterization of the rich critical behaviour encountered in models with
absorbing states is clearly an important issue.  However, lacking a
sound theoretical basis, it is by no means obvious which parameters are
relevant for such a classification.  On general grounds one expects
that the dimensionality $d$ of the system is a relevant parameter.
Moreover, based on the information coming from the one- dimensional
case and from equilibrium one expects that the number of absorbing
states $q$ is another relevant parameter.  Note however that, in some
cases, details of the dynamics might also  change the critical
properties, even without affecting the symmetricity of absorbing
states~\cite{LIPDROZ}. Nevertheless, we can expect that these cases are
accidental rather than generic (as it is the case in equilibrium phase
transitions when a marginal scaling field is present~\cite{BAXTER}).
Keeping in mind the above objections, and lacking a better candidate,
we consider $q$ as a classification parameter~\cite{COMM3}.

In the present paper we study a recently introduced nonequilibrium
Potts model with $q$ absorbing states.  Our numerical results for
$d\geq 2$ can be summarized as follows:  (i) For $d=2$ and $q=3$ the
models has a discontinuous transition.  (ii) For $d=3$ and $q=2$ the
model exhibits a continuous, transition with an order parameter
critical exponent $\beta=1$ (mean-field).  (iii) Simulations are
inconclusive, however, in the $d=2$ and $q=2$ case, and we suggest that
in this case the model is close to or at the crossing point of lines
separating three different types of phase transitions.  Our results
together with the already accumulated knowledge, prompted us to
partition the ($q,d$) plane into three regions of different phase
transitions: mean-field,  non-mean-field, and discontinuous.  Somewhat
surprisingly, the topology of such a partition is the same as for the
equilibrium Potts model.  Although nonequilibrium systems, and in
particular models with absorbing states, are usually regarded as very
much different from equilibrium systems  our work shows, however, that
despite some differences there are also some qualitative similarities.

In addition, we performed simulations of a $d=2$ BARW model without
parity conservation.  Our results confirm the field-theory predictions
according to which the critical behavior belongs to to the directed
percolation universality class.

\section{Nonequilibrium Potts model}

Before presenting our model let us recall basic properties of the
equilibrium Potts model.  First we assign with a lattice site $i$ a
$q$-state variable $\sigma_i=0,1,...,q-1$.  Next, we define the energy
of this model through the Hamiltonian:
\begin{equation}
H= -\sum_{(i,j)} \delta_{\sigma_i\sigma_j},
\label{e1}
\end{equation}
where summation is over pairs $(i,j)$ which are usually nearest
neighbours and $\delta$ is the Kronecker delta function.  
This equilibrium statistical mechanics model was studied
using many different analytical and numerical methods and is a rich
source of the information about phase transitions and critical
phenomena~\cite{WU}.

To study the equilibrium Potts model using Monte Carlo simulations one
constructs a stochastic Markov process with suitably chosen transition
rates.  One of the possible choices corresponds to the so-called
Metropolis algorithm.  In this algorithm one looks at the energy
difference $\Delta E$ between the final and initial configuration and
accept the move with probability min$\{1,{\rm e}^{-\Delta E/T}\}$,
where $T$ is temperature.

A nonequilibrium Potts model having $q$ adsorbing states can be
obtained by making the following transformation in the Metropolis
dynamics~\cite{LIPDROZ}:  when all neighbours of a
given site are in the same  state as this site, then this site cannot
change its state (at least until one of its neighbours is changed).
Let us notice that any of $q$ ground states  of the equilibrium Potts
model (\ref{e1}) is an absorbing state of our nonequilibrium Potts
model. 
The one-dimensional version of this nonequilibrium Potts model has already been examined~\cite{LIPDROZ}.
In addition to recovering the expected critical behaviour for $q=2$ and 3 it was 
found that certain additional modification of its dynamics might affect the critical 
behaviour. We shall not be concerned with such a variant in the present paper.

Since the dynamics of our models is obtained from a modification of the Metropolis
algorithm of an equilibrium system, transition probabilities are parametrized
by temperature-like quantity $T$.
Strictly speaking, for our model the ordinary temperature cannot be defined.
Nevertheless, we will refer to this quantity as temperature.

Of course in the realm of nonequilibrium systems there are also other models
than those with absorbing states.
One of the important questions is whether under certain conditions
nonequilibrium systems might be mapped, at least at the coarse-grained level, 
into equilibrium ones.
Some aspects of this problem were studied by Grinstein et 
al.~\cite{GRINSTEIN}.
In a class of models studied by them it is important that all transition 
probabilities are strictly greater than zero.
In models with absorbing states this requirement is clearly violated.

\section{Monte Carlo simulations and results}

To study the properties of our model we made extensive Monte-Carlo
simulations. A natural characteristic of models with absorbing states
is the steady-state density of active sites $\rho$.
A given site $i$ is active when at least one of its 
neighbours is in a state different than $i$.
Otherwise the site $i$ is called nonactive.
Upon approaching a critical point $\rho$ develops a power-law singularity
characterized by the exponent $\beta$.
At the first-order transition $\rho$ has a jump.
In addition, we also looked at its time dependence
$\rho(t)$.  
In the active phase $\rho(t)$ converges to the positive
value, while at criticality $\rho(t)$ usually has a power-law decay
$\rho\sim t^{-\delta}$.
In the absorbing phase $\rho$ very often decays faster than the power law, 
however, in some cases (also those studied in the present paper) a power law 
behaviour is seen, but with a different exponent than for the critical decay.

Moreover, we used the so-called dynamic Monte Carlo method where one
sets the system in the absorbing state with activity only locally
initiated and monitor some stochastic properties of runs~\cite{GRASSTORRE}.  
One of the most frequently used
characteristics is the survival probability $P(t)$ that activity
survives at least until time $t$ and the average number of active sites $N(t)$
(to calculate $N(t)$ we average over all runs).
At criticality $P(t)$ and $N(t)$ are expected to
have power-law decay: $P(t)\sim t^{-\delta'}$ and $N(t)\sim t^{\eta}$.
(For some models $\delta=\delta'$, but exceptions from this relation are also
known~\cite{HAYE1}).  Our simulations were made for various system
sizes and we ensured that the system was large enough so that presented
below results are size independent.

\subsection*{$\mathbf{d=2,\ q=3}$}

\begin{figure}
\begin{center}
\centerline{\epsfxsize=9cm \epsfbox{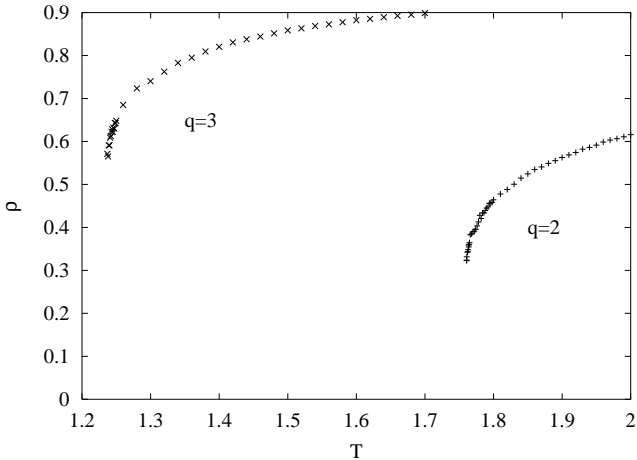}}
\end{center}
\caption{
The density of particles $\rho$ as a function of temperature $T$ for
the $d=2$ nonequilibrium Potts model.
Close to the transitions we used the lattice of the linear size $L=600$.}
\label{ro}
\end{figure}

\begin{figure}
\centerline{\epsfxsize=9cm \epsfbox{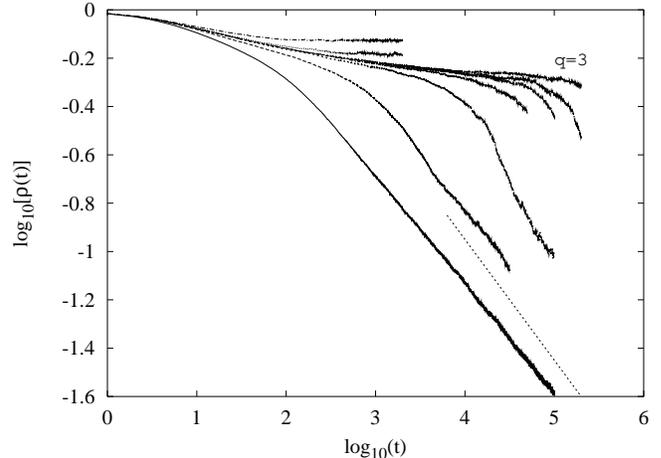}} 
\caption{
The time dependence of the density $\rho(t)$ for the $d=2$
model and (from
top)
$T=$1.3, 1.25, 1.237, 1.2365, 1.236, 1.235, 1.23, 1.2, and 1.1 
($L=500$).
Each line is an average of 100 independent runs which starts from 
random initial 
configurations.
The dotted straight line has a slope corresponding to $\delta=0.5$.
}
\label{t3}
\end{figure}

\begin{figure}
\begin{center}
\centerline{\epsfxsize=9cm \epsfbox{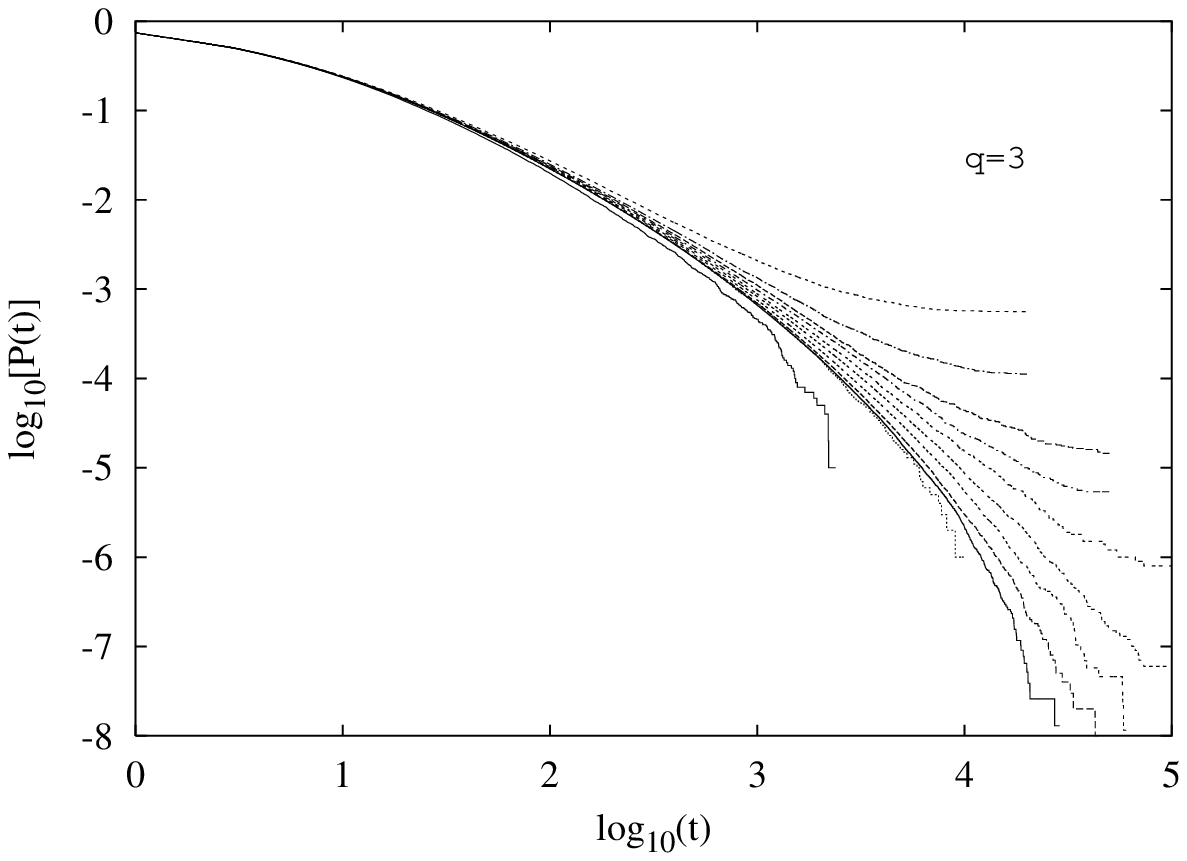}}
\end{center}
\caption{
The time dependence of the survival probability $P(t)$ for the 
$d=2$ model and (from
top) $T=$1.25, 1.245, 1.242, 1.241, 1.24, 1.239, 1.238, 1.237, 
1.2365, 1.236, and 1.23 
($L=500$).
Lines for $T=1.2365$, 1.237, and 1.238 are obtained averaging over 
$10^8$ independent 
runs.
}
\label{dyn3}
\end{figure}

\begin{figure}
\begin{center}
\centerline{\epsfxsize=9cm \epsfbox{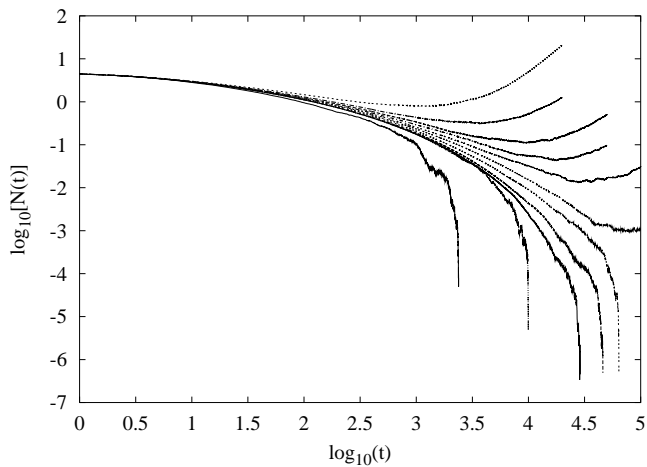}}
\end{center}
\caption{
The time dependence of the number of active sites $N(t)$ for the 
$d=2,\ q=3$ model and (from
top) $T=$1.25, 1.245, 1.242, 1.241, 1.24, 1.239, 1.238, 1.237, 
1.2365, 1.236, and 1.23 
($L=500$).
Lines for $T=1.2365$, 1.237, and 1.238 are obtained averaging over 
$10^8$ independent 
runs.
}
\label{dyn3a}
\end{figure}

Simulations for $d=2$ models were performed on the square lattice.  The
temperature dependence of the density of active sites $\rho$ is shown
in Fig.~\ref{ro}.  For $T<1.237$ it is
virtually impossible to reach an active steady state value of $\rho$, which
suggests that the model undergoes a discontinuous phase transitions.
Such a scenario is confirmed by the data in
Figs.~\ref{t3}-\ref{dyn3a}.  In Fig.~\ref{t3} one can see that upon
approaching the transition point $\rho(t)$ develops a longer and longer
plateau.  At the transition point, which we locate around $T=1.237$ the
density $\rho(t)$ has basically an infinitely-long plateau.  Such a
behaviour is a clear indication of the discontinuous nature of the
transition.
Let us notice that below the transition temperature $\rho(t)$ decays as
$t^{-1/2}$. 
A simple scaling argument can be used to show that such a behaviour is related
with the average domain size growth $l\sim t^{1/2}$.
Indeed, let us consider the $d=2$ system of the linear size 
$L$.
It contains $(\frac{L}{l})^2$ subdomains of the linear size $l$ and thus 
the total perimeter of these subdomains scales as $\frac{L^2}{l}$.
Since active sites are located mainly at the domain walls their density scales as
$\frac{L^2}{lL^2}=l^{-1}$.
Assuming now that $l$ increases as $t^{1/2}$ we obtain that $\rho\sim t^{-1/2}$.
The characteristic length which increases as $t^{1/2}$ is typical for 
coarsening in
the broken-symmetry phase of the original Potts model~\cite{BRAY,GUNTON}.
It also appears in the evolution of one-dimensional nonequilibrium systems which
belong to the parity-conserving universality class.
Results of the dynamic Monte Carlo also support the discontinuous
nature of the transition.  Indeed, both for $P(t)$ (Fig.~\ref{dyn3})
and $N(t)$ (Fig.~\ref{dyn3a}) the data are systematically bending and
no clear power-law behaviour is observed.

Thus, as a  summary, for the $d=2, \ q=3$ case our model undergoes 
a discontinuous phase transition.

\subsection*{$\mathbf{d=2,\ q=2}$}

\begin{figure}
\centerline{\epsfxsize=9cm \epsfbox{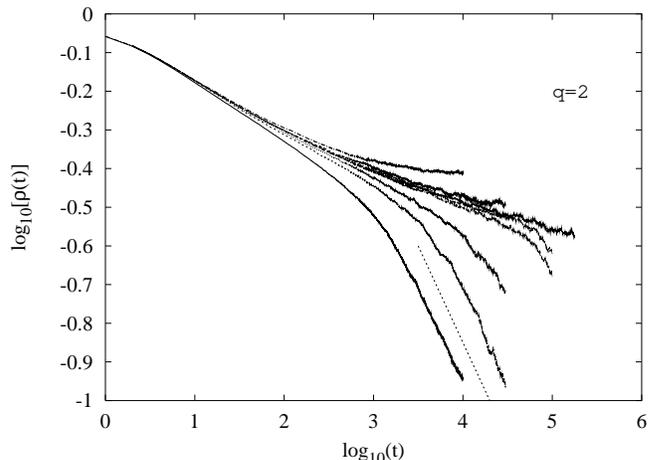}}
\caption{
The time dependence of the density $\rho(t)$ for the $d=2$
model and (from
top)
$T=$1.77, 1.76, 1.7585, 1.758, 1.757, 1.75, 1.74, and 1.71 
($L=500$).
Each line is an average of 100 independent runs which starts from 
random initial configurations.
The dotted straight line has a slope corresponding to $\delta=0.5$.
}
\label{t2}
\end{figure}

\begin{figure}
\centerline{\epsfxsize=9cm \epsfbox{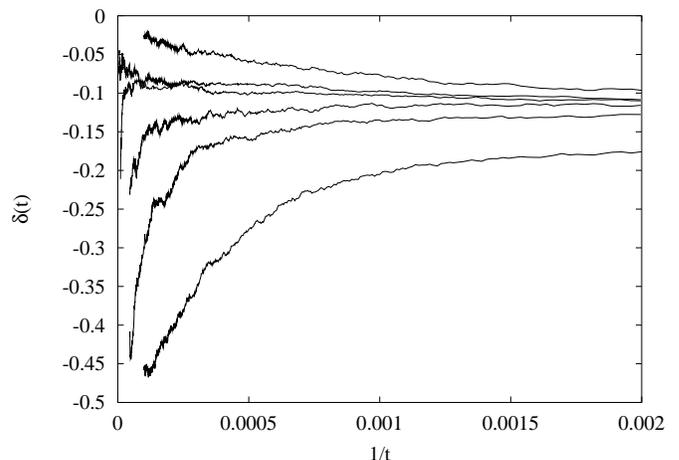}}
\caption{
The exponent $\delta(t)$ as a function of $1/t$ for the $d=q=2$
model and (from
top) $T=$1.77, 1.7585, 1.757, 1.75, 1.74 and 1.71 ($L=500$).
Each line is an average of about $100$ independent runs.
}
\label{flocal}
\end{figure}

In the $q=2$ case the density of active sites is a similar function of
temperature as in the $q=3$ case (Fig.~\ref{ro}).  Let us notice
however, that now the jump in $\rho$ is almost twice smaller than
previously.  We should be also aware of the fact that an observed jump
might be a finite-size effect.  (Large fluctuations might drive the
system into an absorbing state even though temperature is above the
critical temperature).  As further results show, it is rather difficult
to clarify the nature of the transition in this case.  First, let us
notice that time dependence of $\rho(t)$ does not develop a clear
plateau as in the $q=3$ case.  (We estimate the critical point in this
case as $T_c=1.7585(3)$).  But neither there is a pronounced power-law
behaviour seen in Fig.~\ref{t2}.  If $\rho(t)$ does decay as
$t^{-\delta'}$ at criticality then $\delta'$ is very small
($\delta'=0.07(2)$), which suggests that the true exponent $\delta'$
might be equal to zero.  
Sometimes to improve the estimation of $\delta$ one can use the so-called
local slopes method~\cite{HAYE1}.
In this method $\delta$ is calculated from the equation
\begin{equation}
\delta(t)=\frac{{\rm log}_{10}[\frac{\rho(t)}{\rho(t/m)}]}{{\rm log}_{10}(m)},
\label{localslopes}
\end{equation}
where $m$ is a certain constant.
Application of this method to our data and $m=5$ is shown in Fig.~\ref{flocal}.
At criticality the data converge (with some scattering) to the value 0.06 
but such a small value means that the possibility $\delta=0$ still cannot 
be excluded.

\begin{figure}
\centerline{\epsfxsize=9cm \epsfbox{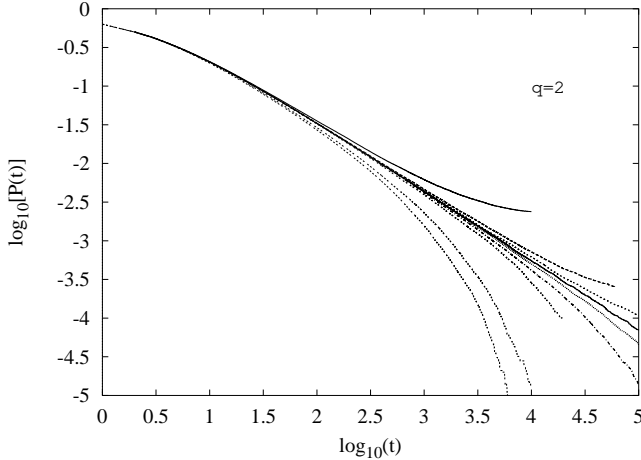}}
\caption{
The time dependence of the survival probability $P(t)$ for the 
$d=2$
model and (from
top) $T=$1.77, 1.76, 1.759, 1.7585, 1.758, 1.757, 1.755, 1.74, and 
1.73 ($L=500$).
Each line is an average of about $10^6$ independent runs.
}
\label{dyn2}
\end{figure}

In our opinion, from the steady-state and
time-dependent measurements of $\rho$ the most likely possibility is
that the transition is of first order and the plateau will develop (in
Fig.~\ref{t2}) but only at a larger time scale.
However, the dynamic Monte Carlo results shed some doubts on such an 
interpretation.
Indeed, in Fig.~\ref{dyn2} we can see that $P(t)$ has a clear power-law decay
for at least three decades in time with the exponent $\delta'=0.90(2)$.
In addition, $N(t)$ seems to remain constant at the transition point which
suggest that $\eta=0$.
To support our dynamic Monte Carlo results let us notice that Hinrichsen
reported basically the same values for dynamical exponents for another
two-dimensional model with $q=2$~\cite{HAYE97}.
\begin{figure}
\centerline{\epsfxsize=9cm \epsfbox{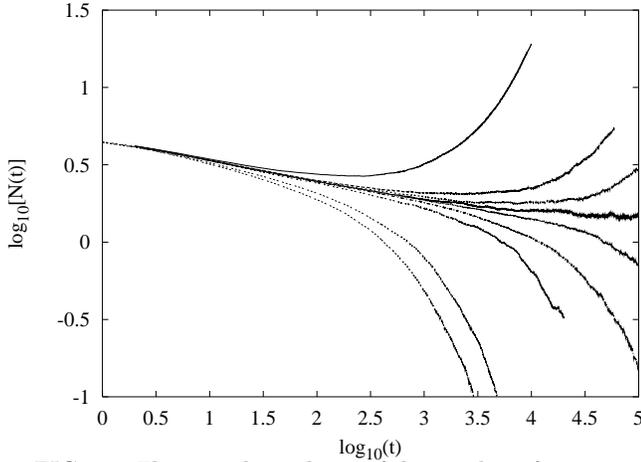}}
\caption{
The time dependence of the number of active sites $N(t)$ for the 
$d=2,\ q=2$
model and (from
top) $T=$1.77, 1.76, 1.759, 1.7585, 1.758, 1.757, 1.755, 1.74, and 
1.73 ($L=500$).
Each line is an average of about $10^6$ independent runs.
}
\label{dyn2a}
\end{figure}
The fact $\delta'$ is close to unity and $\eta=0$ prompted him to suggest that
the model exhibit a mean field behaviour.
In such a case, however, one should have $\delta=1$, which is clearly 
in contradiction with the behaviour of $\rho(t)$ which we observe.
A power-law behaviour of $P(t)$ and $N(t)$ is typical for continuous 
transitions which seems to be in conflict with the behaviour of $\rho$.
However, the possibility that we have a discontinuous transition accompanied
by some dynamical power-law characteristics, although exotic at first sight, 
cannot be ruled out.
We will return to this problem in the next section.

Let us also notice (Fig.~\ref{t2}) that, similarly to the $q=3$ case, 
for $T<T_c$ the density $\rho(t)$ seems to decay in time as $t^{-1/2}$.
\subsection*{$\mathbf{d=3,\ q=2}$}
\begin{figure}
\begin{center}
\centerline{\epsfxsize=9cm \epsfbox{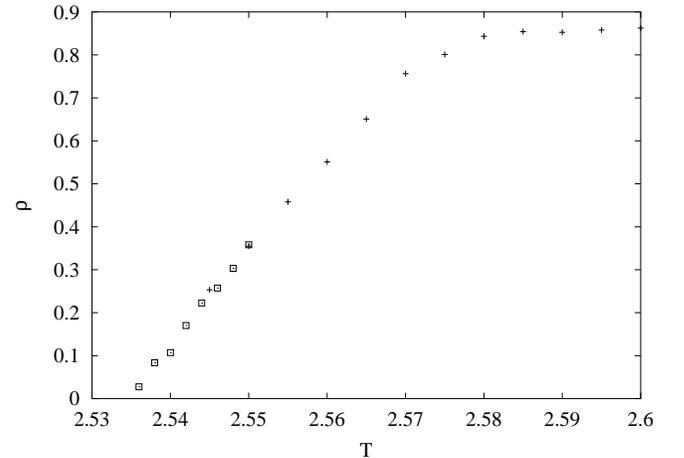}}
\end{center}
\caption{
The density of particles $\rho$ as a function of temperature $T$ for
the three-dimensional $q=2$ nonequilibrium Potts model with $L=60 (\Box)$ and 40(+).
}
\label{ro_cubic}
\end{figure}

As a last case in this section, we consider our model on the simple cubic 
lattice and for $q=2$.
Of course, simulations for three-dimensional systems are very demanding.
Therefore we were not able to perform detailed time-dependent simulations
nor the dynamic Monte Carlo.
However, the steady-state density $\rho$ as measured for $L=40$ and 60
is basically size independent (Fig.~\ref{ro_cubic}).
Nearly linear behaviour of $\rho$ in the vicinity of the transition suggests 
that in this case $\beta=1$, which indicate a mean-field nature of the 
transition.
\subsection*{(q,d) phase diagram}
\begin{figure}
\begin{center}
\centerline{\epsfxsize=9cm \epsfbox{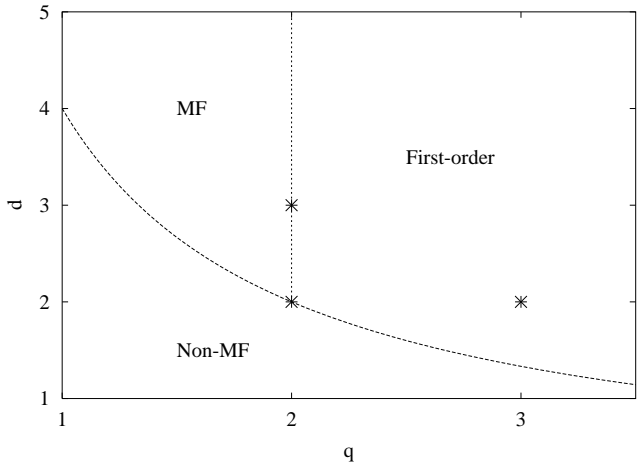}}
\end{center}
\caption{
The location of various types of phase transitions for $d-$dimensional models
with $q$ absorbing states.
Models for which numerical results are reported in this paper are denoted
as stars.
}
\label{diagram}
\end{figure}

In this section we sketch the overall behaviour of our model in the $(q,d)$
plane.
Our proposal (Fig.~\ref{diagram}) is based on the already accumulated knowledge, 
presented above results, and a minimalistic assumption that the resulting 
phase diagram should not be too complicated.
Essentially, we suggest that the $(q,d)$ plane can be divided into three 
regions of different types of phase transitions: (i) non-mean-field, (ii)
mean-field, and (iii) first-order.
Some comments are in order. 
As it is defined only through dynamical rules, our model is trivial
for $q=1$.
However, the $q=1$ case corresponds to  models with a single absorbing state
(directed percolation, contact process)
It is known that the critical dimension $d_c$ in this case equals 4 and
for $d$ smaller/larger than $d_c$ we have a continuous
non-mean-field/mean-field transition.  
Firmly established are also
results along the $d=1$ line.  It is known that $q=2$ case corresponds
typically to the PC universality class.  For $q\geq 3$ one
expects~\cite{HAYE97,CARLON} that the model typically belongs to the
same universality class as a $N$-BARW model studied by T\"auber and 
Cardy~\cite{TAUBER}, although, under
more restrictive dynamics the PC criticality might also appear~\cite{LIPDROZ}.
In any case, the critical behaviour is  non-mean-field.  Taking into
account the mean-field behavior obtained in the case $d=3,\ q=2$, we
assume that this case falls into the same region as DP above critical
dimension.  On the other hand, the discontinuous transition in the
$d=2,\ q=3$ case implies the existence of the third region.  Actually
we performed some simulations also for $d=2,\ q=4$ case.  Basically the
behaviour is similar to the $q=3$ case except that the jump is larger
and the first-order character of the transition is even more
transparent.  We expect that such a behaviour persists for all $q>3$.
Provided that for $d>2$ there are no qualitative changes, the diagram
must have a structure as shown in Fig.~\ref{diagram}.

It is thus clear that our most difficult case $d=q=2$ is located
somewhere close to the point where all three regions meet.  We have no
strong arguments to locate this point exactly at the crossing point, but
such a location would certainly explain unusual behaviour as seen in
our Monte Carlo simulations.  We do not exclude, however, the possibility
that $d=q=2$ case is off the crossing point but somewhere close to it,
which would still explain the numerical difficulties in this case.

The resulting diagram looks very similar to the diagram of the
equilibrium Potts model~\cite{WU}.  In the equilibrium model the
non-mean-field part is basically shifted by two upwards (critical
dimensions for ordinary percolation and the Ising model are 6 and 4,
respectively).  However, the diagram for the equilibrium Potts model is
much more meaningful than in our case.  Indeed, in the
Fortuin-Kasteleyn representation, the Potts model is well defined for
any, even non-integer $q$, which justifies continuous lines on its
phase diagram.  In our case, we do not have a representation of our
model with continuous $q$.  The key property that would be required for
such a representation is the existence of an analog of the partition
function.  Such a quantity exists for nonequilibrium systems only in
very special cases~\cite{DERRIDA}.  Provided that a certain partition
function exists for our model (no matter how complicated), and that
using this function one can find a corresponding Fortuin-Kasteleyn
representation (no matter how complicated), continuous lines in our
diagram would be meaningful.

Of course, the presented diagram does not encompass all models with absorbing 
states.
It is well known that for example there are $d=2,\,q=1$ models with first-order
transitions~\cite{HAYE1}.
But as we already mentioned, similar situation occurs for equilibrium 
systems where certain factors (anisotropies, additional interactions etc.) 
might change more generic behaviour.
The presented diagram is valid only for the presented Potts model and 
its applicability to other systems requires additional examination.
\section{Parity-non conserving BARW model in two dimensions}
\begin{figure}
\begin{center}
\centerline{\epsfxsize=9cm \epsfbox{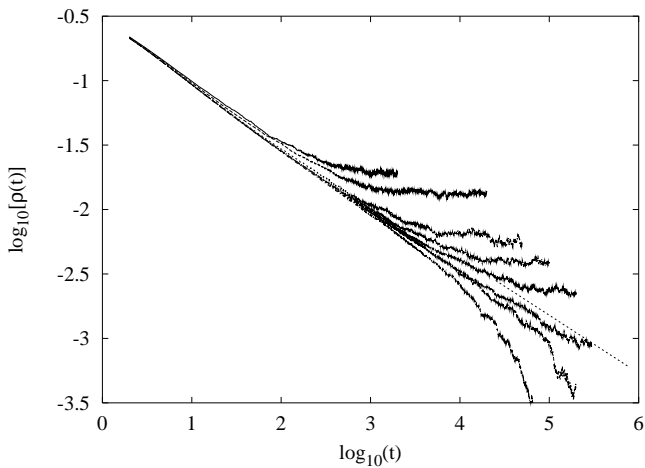}}
\end{center}
\caption{
The time dependence of density of particles $\rho(t)$ for the two-dimensional 
BARW model.
Simulations were made for (from top) $p=$0.8, 0.81, 0.82, 0.822, 0.823, 0.8237, 
0.824, and 0.825.
The dotted line has a slope corresponding to the DP value $\delta=0.451$.
}
\label{time}
\end{figure}

\begin{figure}
\begin{center}
\centerline{\epsfxsize=9cm \epsfbox{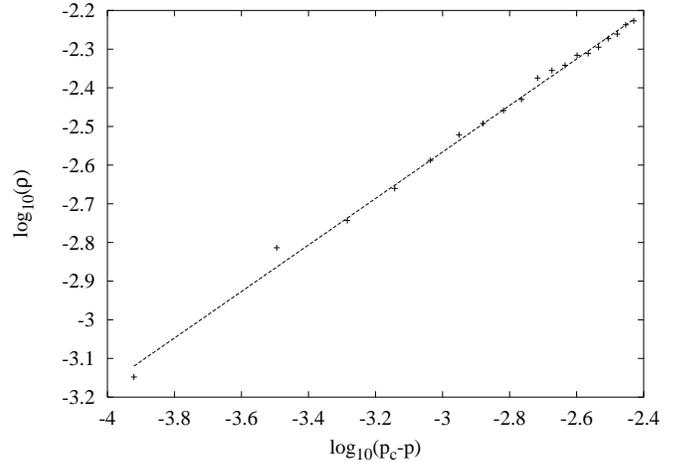}}
\end{center}
\caption{
The density of particles $\rho(t)$ as a function of $p_c-p$ for the 
two-dimensional BARW model ($p_c=0.8237$).
Only results for $L=2000$ runs are shown here.
The linear fit (dotted line) obtained using the least-square method 
has a slope corresponding to $\beta=0.6$.
}
\label{ro_barw_log}
\end{figure}

In the present section we examine the two-dimensional BARW model
without parity conservation.  In this model particles are located on
sites of a square lattice.  In addition to diffusion, which takes place
at the rate $p$, particles can branch, at the rate $1-p$, according to
the following reaction:
\begin{equation}
X\rightarrow 2X,
\label{x2x}
\end{equation}
where the offspring particle is placed on the randomly chosen nearest
neighbour of a parent particle.  Moreover, two particles which happen
to be placed at the same site annihilate instantaneously
\begin{equation}
2X\rightarrow 0
\label{2x0}
\end{equation}
This model was already examined by Takayasu and Tretyakov~\cite{TT}.
They suggested that the model undergoes a continuous transition around
$p=0.85$ and the density of particles decays linearly at the transition
point ($\beta=1$).  This result contradicts more recent field-theory
approach which suggests that in this case the model should belong to
the DP universality class~\cite{TAUBER}.
\begin{figure}
\begin{center}
\centerline{\epsfxsize=9cm \epsfbox{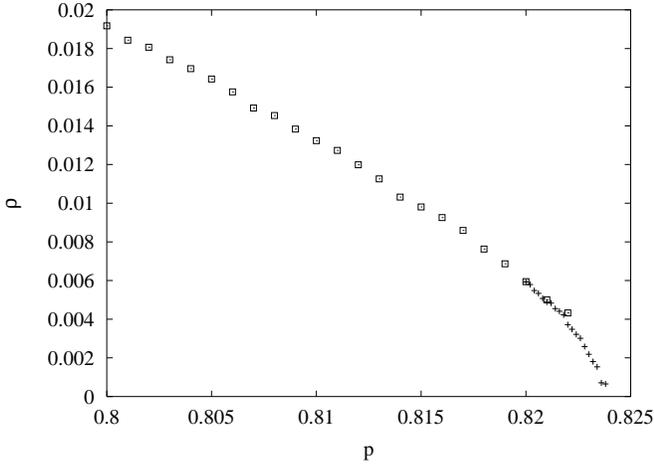}}
\end{center}
\caption{
The density of particles $\rho$ as a function of the diffusion rate $p$ for
the two-dimensional BARW model with $L=1000 (\Box)$ and 2000(+).
}
\label{ro_barw}
\end{figure}
Since such a disagreement requires an explanation, we performed Monte Carlo 
simulations of this model.
Our system size was much larger than in Takayasu and Tretyakov simulations
and we approached much closer to the critical point.
One can see (Fig.~\ref{ro_barw}) that although around $p=0.8$ (which was the
largest value of $p$ simulated by Takayasu and Tretyakov) the density seems to 
decay linearly it has a pronounced bending close to the transition point.
To obtain more accurate estimation of the critical point we examined the time
dependence of $\rho(t)$ (see Fig.~\ref{time}).
From these analysis we obtain the following estimation of the critical point
$p_c=0.8237(5)$.
One can also see that at criticality $\rho(t)$ has a power-law decay with
the exponent close to the DP value $\delta=0.451$.
Having the critical point we can estimate exponent $\beta$ and the 
corresponding data are shown in Fig.~\ref{ro_barw_log}.
The least-square fit gives $\beta=0.60(3)$ which is certainly compatible with
the DP value 0.584(4).
To summarize this section, our results confirm the field-theory 
prediction that BARW models without parity conservation belong to the DP
universality class.
\section{Conclusions}
In the present paper we examined $d$-dimensional nonequilibrium models with $q$ absorbing  states.
As our main result we obtained the diagram shown in Fig.~\ref{diagram}.
Interestingly, this diagram bears some similarity to the diagram of equilibrium 
Potts model.
In addition, we clarified the nature of the phase transition in the $d=2$ 
BARW model without parity conservation.
Together with the work of Szab\'o and Santos~\cite{SZABO} 
for the parity conserving case,
it confirms predictions of the field-theory for $d=2$ BARW models by Cardy and 
T\"auber~\cite{TAUBER}.

Although it requires considerable numerical efforts, it 
would be desirable to clarify the behaviour of the $d=q=2$ model.
Our results are inconclusive in this case, but a possibility that an 
interesting critical behaviour could be found should motivate
further study.
\acknowledgements
This work was partially supported by the Swiss National Science Foundation
and the project OFES 00-0578 "COSYC OF SENS".


\begin{references}
\bibitem{COMM2} In the surface-catalysis models active sites very often 
constitute boundaries of inactive domains, i.e., they are located on 
$d-1$ dimensional manifolds.
To consider such sites as particles we would have to use a number of 
constraints, which would make a resulting BARW model extremely
complicated.
However, in the $d=1$ models domain boundaries are point-like and relations
with BARW models hold.
\bibitem{INFINITE} In the present paper we do not consider models with 
infinitely many absorbing states.
In some cases typical universality classes are reported like DP or PC but 
it is also possible that some models of this kind might exhibit novel critical
behaviour.
Classification of such models into universality classes is certainly more 
difficult than of models with finitely many absorbing states, which are of 
primary concern in the present paper.
\end{references}
\end{document}